\documentclass[pre,preprint, superscriptaddress, showpacs,amsmath,amssymb]{revtex4}
\usepackage{graphicx}

\begin{document}

\title{Linear dielectric response of clustered living cells}

\author{Titus Sandu}
\affiliation{International Center for Biodynamics, Bucharest, Romania}
\email{egheorghiu@biodyn.ro}
\author{Daniel Vrinceanu}
\affiliation{Department of Physics, Texas Southern University, Houston, Texas 77004, USA}
\author{Eugen Gheorghiu}
\affiliation{International Center for Biodynamics, Bucharest, Romania}

\date{\today}

\begin{abstract}
The dielectric behavior of a linear cluster of two or more living cells connected by tight junctions is analyzed using a 
spectral method. The polarizability of this system is obtained as an expansion over the eigenmodes of the linear
response operator, showing a clear separation of geometry from electric parameters. The eigenmode with the
second largest eigenvalue dominates the expansion as the junction between particles tightens, but only 
when the applied field is aligned with the cluster axis. This effect explains a distinct low-frequency
relaxation observed in the impedance spectrum of a suspension of linear clusters.

\end{abstract}

\maketitle
\section{Introduction}

Particle polarizability governs the electric response for many inhomogeneous systems ranging from biological cells
to plasmonic nanoparticles and depends strongly on both its dielectric and geometric properties.
Analytical models have been reported \cite{Pauly1959, Asami1980} only for spherical and ellipsoidal geometries,
whereas more complex geometries have been approached by direct numerical solution of the field equations
using, for example, the finite difference methods \cite{Asami2006}, the finite element method \cite{Fear1998},
the boundary element method \cite{Sekine2005,Sancho2003}, or the boundary integral equation (BIE) \cite{Brosseau2003}. 

In a simplified representation, biological cells can be regarded as homogeneous particles (cores)
covered by thin membranes (shells) of contrasting electric conductivities and permittivities.
Complex geometries occur when cells are undergoing division cycles (e.g. budding yeasts) or are 
coupled in functional tissues (e.g. lining epithelia or myocardial syncytia).
In these cases, the dielectric/impedance analysis of cellular systems is far more complicated than
previous models 
\cite{Schwan1996,Gheorghiu1993,Gheorghiu1994},
which considered suspensions of spherical particles.
Intriguing dielectric spectra \cite{Knapp2005} 
reveal distinct dielectric dispersions with time evolutions consistently related to
tissue functioning or alteration, identifying a possible role of cell connectors
(gap junctions) in shaping the overall dielectric response.

A direct relation between the microscopic parameters and experimental data can be analytically 
derived only for dilute suspensions of particles of simple shapes, and is rather challenging for
system with more realistic shapes,
where only purely numerical solutions have been available.
In this work we demonstrate that a spectral representation of a BIE 
provides the analytical structure for the polarizability of particles with a wide range of shapes and structures. 
The numerically calculated parameters encode particle's geometry information and are accessible by experiments. 

By using single and double-layer potentials \cite{Vladimirov1984}, the Laplace equation for the fields inside
and outside the particle is transformed into an integral equation. A spectral representation
for the solution of this equation is obtained providing the eigenvalue problem for the linear
response operator is solved. Although not symmetric, this operator has a real spectrum bounded by
-1/2 and 1/2 \cite{Ouyang1989,Fredkin2003,Mayergoyz2005} and its eigenvectors are orthogonal to those of the
conjugate double-layer operator. A matrix representation is obtained by using a finite basis of surface 
functions.

The true advantage of the spectral method is that the eigenvalues and eigenvectors 
of the integral operator  provide valuable insight into the dielectric behavior of clusters of biological cells.
The eigenvectors are a measure of surface charge distributions due to a field.
Only eigenvectors with a non-zero dipole moment contribute to the polarizability of the particle.
We call these dipole-active eigenmodes. An effective separation of the geometric and morphologic properties
from  dielectric properties is therefore achieved \cite{Vrinceanu1996}.
We also show that for a particle covered by multiple confocal shells,
the relaxation spectrum is a sum of Debye terms with the number of relaxations equal to the number 
of interfaces times the number of dipole-active eigenvalues.
This is a generalization of a previous result \cite{Hanai1988} on cells of arbitrary shape. 

Our method is related to another spectral approach which uses an eigenvalue differential equation
\cite{Bergman1978,Bergman1992,Stockman2001,Li2003}. This method has been applied to
biological problems by Lei \textit{et al.} \cite{Lei2001} and by Huang \textit{et al.} \cite{Huang2002}. 
These authors, however, considered homogeneous cells with much simpler expression for cell polarizability.
The BIE spectral method seeks a solution on the boundary surface defining the particle, as opposed to
the eigenvalue differential equation, where the solution  is defined in the entire space.

In a previous study on double (budding) cells it was shown that before cells separation an 
additional dispersion occurs \cite{Asami1998}.
Moreover, in recent papers \cite{Asami2006,Gheorghiu2002} numerical experiments have shown that the dielectric 
spectra of a suspension of dimer cells connected by tight junctions exhibit an additional, distinct low-frequency 
relaxation.
Our numerical calculation shows that the largest dipole-active eigenvalue approaches the value of 1/2 as
the junction become tighter.
Although the coupling of this eigenmode with the electric field stimulus is relatively modest
(the coupling weight is about 1-2 \%), this eigenmode 
has a significant contribution to the polarizability of clusters.
Thus the eigenmodes close to 1/2 induce an additional low-frequency 
relaxation in the dielectric spectra of clustered biological cells even though the coupling is quite small.
Needle-like objects, such as elongated spheroids or long cylinders, have similar polarizability features.

In this paper we consider rotationally symmetric linear clusters made of up to 4 identical particles 
covered by thin insulating membranes and connected by
junctions of variable tightness.
Convenient and flexible representations for the surfaces describing these objects
are provided.
The number of relaxations in the dielectric spectrum of the linear clusters,
their  time constants and their relative strengths are analyzed in terms of the eigenmodes of the
linear response operator specific to the given shape.

\section{Theory}

\subsection{Effective permittivity of a suspension}

We consider a suspension of identical, randomly oriented particles of arbitrary shape and
dielectric permittivity, $\varepsilon _{1} $, immersed 
in a dielectric medium of dielectric permittivity, $\varepsilon _{0} $.
The dielectric permittivities
are in general complex quantities and the theory described here applies also for time-dependent fields, providing that
the size of a particle is much smaller than the wavelength. 
When an applied uniform electric field interacts with the suspension,
the response of the system is linear with the applied field and an
effective permittivity for the whole sample can be measured and is defined by
\cite{Gheorghiu2002,Jackson1975,Prodan1999}:
\begin{equation} \label{eq2} 
\varepsilon_{\mbox{\tiny sus}} =\varepsilon_0 + f\frac{\alpha \varepsilon_0 }{1-f\frac{\alpha }{3} } . 
\end{equation} 
This result is obtained in the limit of low concentration, weak intensity of the stimulus field, and
using an effective medium theory within the dipole approximation.
Here $f = N V_1/V$ is the volume fraction of all $N$ particles, each of volume $V_1$, with respect to the 
total volume of the suspension $V$. The averaged normalized polarizability $\alpha$ of a particle is defined as 
\cite{Prodan1999,Gheorghiu2002,Sebastian2008}
\begin{equation} \label{eq1} 
\alpha =\frac{1}{4\pi {\kern 1pt} V_{1} } \int _{V_{1} } \int _{\Omega _{N} } 
\left(\frac {\varepsilon _{1} -\varepsilon _{0}}{\varepsilon _{0}} 
\right){\bf E}\left({\bf N}\right) \cdot {\bf N}\;d\Omega _{N}\;dV 
\end{equation}
where ${\bf E}\left({\bf N}\right)$ is the electric field perturbation created inside the particle under a
normalized applied electric excitation with direction ${\bf N}$ and $d\Omega _{N} $ 
is the solid angle element generated by that direction. The above normalized polarizability is 
dimensionless and is obtained by multiplying the standard polarizability of a particle with the 
factor $4 \pi/V_1$. In the following we will refer only to normalized polarizability, 
thus, without any confusion, the normalized polarizability $\alpha$ will be simply called polarizability. 
The directional average in \eqref{eq1}  is equivalent to the averaged sum 
over three orthogonal axes due to the fact that the problem is linear with respect to the applied field. 
The latter is more convenient from computational point of view.

The electric field inside a particle is obtained by solving the following Laplace equation for the
electric potential $\Phi$:
\begin{equation} \label{eq3} 
\begin{array}{rll} 
\Delta \Phi \left( {\bf x}\right) & = 0, & {\bf x} \in \Re ^{3} \backslash \Sigma \\ 
\left. \Phi \right|_{+} & =\left. \Phi \right|_{-} ,  & {\bf x} \in \Sigma  \\ 
\left. {\varepsilon}_{0} \frac{\partial \Phi }{\partial \bf n} \right|_{+} & =
\left. {\varepsilon}_{1} \frac{\partial \Phi }{\partial \bf n} \right|_{-} , \quad & {\bf x} \in \Sigma\\ 
\Phi & \to - {\bf x} \cdot {\bf N} ,  &  \left|{\bf x} \right|\to \infty
\end{array} 
\end{equation} 
where $\Re^{3}$ is the euclidian 3-dimensional space and $\Sigma$ is the surface of the particle.
The derivatives are taken with respect to the normal vector $\bf n$ to the surface $\Sigma$.

Due to the mismatch between the polarization inside and outside the object, electric charges accumulate at the 
interface $\Sigma $ and create an electric potential which counteracts the uniform electric field stimulus. 
The solution of the above Laplace problem (\ref{eq3}) is therefore formally given by 
\begin{equation} \label{eq4} 
\Phi({\bf x})= - {\bf x} \cdot {\bf N}+\frac{1}{{\rm 4\pi }} 
\int _{\Sigma }\frac{ \mu({\bf y})}{|{\bf x} - {\bf y}|} \; d\Sigma ({\bf} y)  
\end{equation} 
The single layer charge distribution $\mu$ induced by the normalized electric field is a solution of the following BIE,
obtained by inserting solution (\ref{eq4}) in equations (\ref{eq3})
\begin{equation} \label{eq5} 
\frac{\mu ({\bf x})}{2\lambda } -\frac{1}{4\pi } \int _{\Sigma }{\rm \mu }({\bf y})\; 
\frac{{\bf n}({\bf x}) \cdot ({\bf x} - {\bf y})}{|{\bf x} - {\bf y}|^3 } \; d\Sigma ({\bf y}) 
= {\bf n} ({\bf x} ) \cdot {\bf N}
\end{equation} 
Here the parameter $\lambda =({\varepsilon}_{1} -{\varepsilon}_{0} )/({\varepsilon}_{1} +{\varepsilon}_{0} )$
isolates all the information regarding the dielectric properties for this problem.

On using the linear response operator $M$ that acts on the Hilbert space of integrable 
functions on the surface $\Sigma$,
\begin{equation} \label{eq6} 
M[\mu ]=\frac{ 1}{4\pi} \int _{\Sigma }{\rm \mu }({\bf y})\;
\frac{{\bf n} ({\bf x}) \cdot ({\bf x} - {\bf y})}{|{\bf x} - {\bf y}|^{3} }\; d\Sigma ({\bf y}) , 
\end{equation} 
the integral equation \eqref{eq5} is written as
\begin{equation} \label{eq7} 
(1/(2\lambda) - M) \mu = {\bf n} \cdot {\bf N} . 
\end{equation} 
The integral operator \eqref{eq6} is the electric field generated by the single layer charge distribution $\mu$ 
along the normal to the surface.  It encodes the geometric information and has several interesting 
properties \cite{Ouyang1989,Fredkin2003,Mayergoyz2005}. Its spectrum is discrete and it is not 
difficult to show that all of its eigenvalues are bounded by the [-1/2, 1/2] interval.
Although non-symmetric, the operator \eqref{eq6} has real non-degenerate eigenvalues.
The eigenvectors are biorthogonal, i. e., they are not orthogonal among themselves,
but orthogonal to the eigenvectors of the adjoint operator 
\begin{equation} \label{eq8} 
M^{\dag} [\mu]=\frac 1{4\pi} \int_{\Sigma } \mu({\bf y})\;
\frac{{\bf n} ({\bf y}) \cdot ({\bf x} - {\bf y})}{|{\bf x} - {\bf y} |^{3} } \; 
d\Sigma ({\bf y}) , 
\end{equation} 
which is associated with the electric field generated by a surface distribution of electric dipoles
(double layer charge distribution).
Therefore, if $|u_{k}\rangle $ is a right eigenvector of $M$ corresponding to eigenvalue $\chi_k$,
$M|u_{k}\rangle =\chi _{k} |u_{k}\rangle$
and
$\langle v_{k'}|$ is a left eigenvector corresponding to eigenvalue $\chi_{k'}$,
$\langle v_{k'}|M^\dag =\chi _{k'} \langle v_{k'} |$, then
\begin{equation} \label{eq10} 
\langle v_{k'} |u_{k} \rangle =\delta _{k'k} , 
\end{equation} 
with the scalar product defined as the integral over the interface $\Sigma $,
\begin{equation} \label{eq11} 
\left\langle f_{1} |f_{2} \right\rangle =\int _{\Sigma }f_{1}^{*} \left({\bf x}\right)f_{2} 
\left({\bf x}\right)d\Sigma ({\bf x}) . 
\end{equation} 

The value 1/2 is always the largest eigenvalue of the operator $M$, regardless the geometry of the object.
This is immediately seen if the object is considered 
to be conductor (${\varepsilon}_{1} \to \infty $), and then the interior electric field has to be zero.
In that case $\lambda =1$, and the charge density that 
generates a vanishing internal electric field obeys the equation $(1/2 - M)\mu=0$,
and therefore 1/2 is an eigenvalue of $M$.
However, this eigenmode is not dipole-active and does not contribute to the total 
polarization of the object.
The operator \eqref{eq6} is insensitive to a scale transformation, which means that its
eigenvalue and eigenvectors depend only on 
the shape of the object and not on its size, or electrical properties.

By employing the spectral representation of the resolvent of the operator $M$
\begin{equation} \label{eq14} 
(z-M)^{-1} =\sum _{k} (z-\chi _{k} )^{-1} |u_{k} \rangle \langle v_{k}|,
\end{equation} 
the solution of equation \eqref{eq7} is obtained for $z=1/(2\lambda )$ as
\begin{equation} \label{eq15} 
{\rm \mu }=\sum _{k} \langle v_k |{\bf n} \cdot {\bf N} 
\rangle ( 1/(2\lambda) - \chi_k )^{-1} |u_k\rangle . 
\end{equation} 

The polarizability of the homogeneous particle is obtained by using the distribution
\eqref{eq15} to build the solution \eqref{eq4} of the Laplace equation  and use it in equation \eqref{eq1}. 
It has been shown that, operationally, the polarizability is 
simply the dipole moment of the distribution \eqref{eq15} over unit volume \cite{Gheorghiu2002,Prodan1999}  

\begin{equation} \label{eq_polarizability} 
\alpha =\frac{1}{3} \frac{1}{V_{1} } \sum _{i,k} \frac{\langle {\bf x} \cdot {\bf N}_i |u_k \rangle
\; \langle v_{k} |{\bf n} \cdot {\bf N}_{i} \rangle }{1/(2\lambda)-\chi_k } , 
\end{equation} 
where ${\bf N}_{i} $ are three mutually orthogonal vectors (directions) of unit norm. 
The factor $(1/(2\lambda)-\chi _{k} )^{-1} $ is a generalized Clausius-Mosotti factor.
Each dipole-active eigenmode contributes to $\alpha $ according to its weight
$p_{k} = \frac{1}{3} \frac{1}{V_{1} }\sum _{i} \langle {\bf x} \cdot {\bf N}_i |u_k \rangle
\; \langle v_{k} |{\bf n} \cdot {\bf N}_{i} \rangle   $,
which determines the strength of coupling between the uniform electric field and
the $k$-th eigenmode and contains three components 
$P_{k,i} =\langle {\bf x} \cdot {\bf {N}_i}|u_{k} \rangle \; \langle v_{k} |{\bf n} \cdot {\bf {N}_i}\rangle /V_{1} $.
Equation \eqref{eq_polarizability} shows a clear separation of the electric properties,
which are included only in $\lambda$, 
from the geometric properties expressed by $\chi_k$ and $p_k$.

\subsection{Shelled particles}
The polarizability of an object covered by a thin shell with permittivity ${\varepsilon}_S $
can be calculated in a similar fashion. The electric field is now 
generated by two single layer distributions, and boundary conditions are imposed twice,
for $\Sigma_1$ and for $\Sigma_2$. The surface $\Sigma_1$ is the outer surface of the shell 
and $\Sigma_2$ is the interface between 
the particle and the shell. The solution of a shelled 
particle in terms of single layer potentials has the form \cite{Sebastian2008}
\begin{widetext}
\begin{equation} \label{eq16} 
\Phi({\bf x})= - {\bf x} \cdot {\bf N} + 
\frac 1{4\pi} \int_{\Sigma_1} \frac{\mu_1({\bf y})}{|{\bf x} - {\bf y}|} \; d\Sigma({\bf y}) + 
\frac 1{4\pi} \int_{\Sigma_2} \frac{\mu_2({\bf y})}{|{\bf x} - {\bf y}|} \; d\Sigma({\bf y}), 
\end{equation} 
\end{widetext}
where $\mu_1$ and $\mu_2$ are the densities defined on surface $\Sigma_1$ and $\Sigma_2$, respectively.
Four integral operators $M_{11}$, $M_{12} $, $M_{21} $ and $M_{22} $ are defined,
depending on which surface are variables ${\bf x}$ and ${\bf y}$. 
For example $M_{11} $ is defined when ${\bf x}$ and ${\bf y}$ are both on $\Sigma _{1} $, 
$M_{12} $ is defined by ${\bf x}$ on $\Sigma _{1} $ and ${\bf y}$ on 
$\Sigma _{2} $, and so on. Thus
\begin{equation} \label{eq17} 
M_{ij}[\mu_j]=\frac 1{4\pi } \int_{\Sigma_j}
\mu_j ({\bf y})\; \frac{{\bf n}({\bf x})\; \cdot  ({\bf x} - {\bf y})}
{|{\bf x} - {\bf y}|^3 }\; d\Sigma({\bf y}) 
\end{equation} 
for $i,j=1,2$. The equations obeyed by $\mu _{1} $ and $\mu _{2} $ are
\begin{equation} \label{eq18} 
\begin{array}{l} 
\mu_1/(2\lambda_1) - M_{11}[\mu_1] - M_{12}[\mu_2] = {\bf n} \cdot {\bf N}\\
\mu_2/(2\lambda_2) - M_{21}[\mu_1] - M_{22}[\mu_2] = {\bf n} \cdot {\bf N} .
\end{array}
\end{equation}
Here the electric parameters are: 
$\lambda_1=(\varepsilon_S - \varepsilon_0)/(\varepsilon_S + \varepsilon_0)$, and 
$\lambda_2=(\varepsilon_1 - \varepsilon_S)/(\varepsilon_1 + \varepsilon_S)$.

We further assume a confocal geometry, i. e. the surface $\Sigma_1$
is a slightly scaled version of $\Sigma_2$,
with a scaling factor $\eta $ close to unity. 
This assumption does not provide constant thickness for the shell,
but our main results should remain at least qualitatively valid \cite{Asami2006,Sancho2003}. 
 
In the limit of very thin shells, and using the scaling properties of the operator $M$, one can show
\cite{Prodan1999,Gheorghiu2002} that all four $M$ operators are related to $M = M_{11}$ 
\begin{equation} \label{eq19} 
\begin{array}{l} 
M_{12}[\mu] = \eta^{-3} ( \mu/2 + M[\mu]) \\ 
M_{12}[\mu] = -\mu/2  + M[\mu] \\ 
M_{22}[\mu] = M[\mu]. \end{array} 
\end{equation} 
Equations \eqref{eq18} can then be arranged in a matrix form as
\begin{equation} \label{eq20} 
\left(\begin{array}{cc}
1/(2\lambda_1 - M) & (1/2 + M)/\eta^3 \\
-1/2 + M & 1/(2\lambda_2) - M
\end{array}\right)
\left(\begin{array}{c} \mu_1 \\ \mu_2 \end{array}\right)
= \left(\begin{array}{c} {\bf n} \cdot {\bf N} \\ {\bf n} \cdot {\bf N} \end{array}\right). 
\end{equation} 
By knowing the eigenvectors and the eigenvalues of $M$ the charge densities $\mu _{1} $ and $\mu _{2} $ 
can be found by inverting the matrix in \eqref{eq20}. For example, $\mu _{1} $ is
\begin{widetext}
\begin{equation} \label{eq21} 
\mu _{1} =\sum _{k} \frac{\eta ^{3} (\frac{1}{2\lambda _{2} } -\chi _{k} )+
\left(\frac{1}{2} +\chi _{k} \right)}{\eta ^{3} \left(\frac{1}{2\lambda _{1} } - 
\chi _{k} \right)\left(\frac{1}{2\lambda _{2} } -\chi _{k} \right)+
\left(\frac{1}{2} +\chi _{k} \right)\left(\frac{1}{2} -\chi _{k} \right)} \langle v_k |{\bf n} \cdot {\bf N} \rangle |u_k\rangle. 
\end{equation} 
\end{widetext}

The field generated by the the two distributions $\mu_1$ and $\mu_2$ outside the particle is the same
as the field generated by an equivalent single layer distribution
\begin{equation} \label{eq25} 
\mu_{e} =\sum _{k} \langle v_{k} |{\bf n} \cdot {\bf N}\rangle (1/(2\tilde{\lambda} _{k} )-\chi _{k} )^{-1} |u_{k} \rangle , 
\end{equation}
where $\tilde{\lambda} _{k} =(\tilde{\varepsilon}_{k} -{\varepsilon}_{0} )/(\tilde{\varepsilon}_{k} +{\varepsilon}_{0})$
and the equivalent permittivity $\tilde{{\varepsilon}}_{k}$ is defined for each eigenmode as: 
\begin{equation} \label{eq26} 
\tilde{\varepsilon}_k ={\varepsilon}_{S} \left(1+\frac{\varepsilon_1 - \varepsilon_S}
{\varepsilon_S + \delta (1/2-\chi_k)\varepsilon_1 +\delta (1/2+\chi_k)\epsilon_S} \right), 
\end{equation}
where $\delta =\eta ^{3} -1 \ll 1$.
The distribution \eqref{eq25} is similar with the distribution \eqref{eq15} obtained for a homogeneous particle,
except that $\lambda $ has to be replaced for each mode with an equivalent 
quantity $\tilde{\lambda} _k$.
Equation \eqref{eq26} can be applied recursively for a multi-shelled structure.
The strict separation of electric and geometric properties is weakened in this case,
because the shape-dependent eigenvalue $\chi_k $ appears now in the 
electric equivalent quantity $\tilde{\lambda} _k $.

The polarizability of the shelled particle is obtained by using the distribution
\eqref{eq25} to build the solution \eqref{eq4} of the Laplace equation  and use it in equation \eqref{eq1},
to get
\begin{equation} \label{eq28} 
\alpha =\frac{1}{3} \frac{1}{V_{1} } \sum _{i,k} \frac{\langle {\bf x} \cdot {\bf N}_i |u_k \rangle
\; \langle v_{k} |{\bf n} \cdot {\bf N}_{i} \rangle }{1/(2\tilde{\lambda}_k)-\chi_k } . 
\end{equation} 
Equation \eqref{eq28} is obtained by replacing $\lambda $ with 
 $\tilde{\lambda} _k $ in Eq. \eqref{eq_polarizability}. The parameter $V_1$ in Eq. \eqref{eq28} 
is the total volume of the cell (the core and the shells).
In the limit of a dilute suspension of identical 
shelled particles, with a low volume fraction $f$,
the effective permittivity \eqref{eq2} is
\begin{equation} \label{eq30} 
\varepsilon_{\mbox{\tiny sus}} = \varepsilon_0 \left(1 + f \sum_k p_k \frac{\tilde{\varepsilon}_k - \varepsilon_0}
{(1/2+\chi_k)\varepsilon_0 + (1/2-\chi_k)\tilde{\varepsilon}_k} \right). 
\end{equation} 

\subsection{The Debye relaxation expansion}

In general, the effective permittivity ${\epsilon}_{{\rm sus}}$ of a suspension of
objects with $m$ shells will have \textit{m}+1 Debye relaxation 
terms for each dipole active eigenmode.
The proof is recursive and 
is based on partial fraction expansion  
with respect to variable $i\omega $ of equations \eqref{eq28} and \eqref{eq30},
provided that the complex permittivity of various dielectric phases 
is ${\epsilon}=\varepsilon -{\rm i\sigma }/(\omega \varepsilon_{vac}) $ where $i=\sqrt{-1} $ and 
$\varepsilon_{vac}$ is the permittivity of the free space ($8.85 \times 10^{-12}$ F/m).
Thus the first Debye term comes out from \eqref{eq30} and 
the remaining $m$ Debye terms result from \eqref{eq26}
by the homogenization process described for shelled particles.
Hence, a suspension of cells with $m$ 
shells (and $m+1$ interfaces) has a dielectric spectrum containing a number of Debye terms equal to $m+1$
times the number of dipole-active eigenvalues. 

The suspension effective permittivity ${\epsilon}_{{\rm sus}}$ has the expansion
\begin{equation} \label{eq42} 
{\epsilon}_{{\rm sus}} ={\epsilon}_{{\rm f}} +\sum _{k,j}\Delta \varepsilon _{kj} /(1+{\rm i\omega T}_{{\rm kj}} )  
\end{equation} 
where ${\epsilon}_{{\rm f}} =\varepsilon _{{\rm hf}} -{\rm i\sigma }_{{\rm lf}} /(\omega \varepsilon_{vac}) $, 
$\varepsilon _{{\rm hf}} $ is the high-frequency permittivity, 
and ${\rm \sigma }_{{\rm lf}} $ is the low-frequency conductivity;
$\Delta \varepsilon _{kj} $ and $T_{{\rm kj}}$ are the dielectric decrement 
and the relaxation time of the \textit{kj} Debye term, respectively;
index $k$ enumerates the dipole-active eigenmodes and index $j$  enumerates interfaces. 

Although the measurable bulk quantities in equation \eqref{eq42} are directly correlated with the microscopic
(electric and shape) parameters, a solution of the inverse problem, which aims at obtaining the microscopic 
information non-intrusively, from the effective permittivity, is in general difficult, if not impossible for the
general multi-shell structure. However, biological cell has a thin and almost non-conductive membrane, and several
simplifications and approximations can be made.
Two Debye relaxation terms in the effective permittivity ${\epsilon}_{{\rm sus}}$ are expected
for each dipole-active eigenmode, corresponding to
the two interfaces which define the membrane.

The first relaxation is derived from the equivalent permittivity \eqref{eq26} which can be written also as Debye 
relaxation terms:
\begin{equation} \label{eq43} 
\tilde{{\epsilon}}_k =\varepsilon +\Delta \varepsilon /(1+ i\omega T ).  
\end{equation} 
The relaxation time $T$ that is given by the poles of $\tilde\epsilon_k$ in \eqref{eq26} is a quite good approximation 
of the first relaxation time $T_{k1}$
\begin{equation} \label{eq44} 
T = \varepsilon_{vac} \frac{\left(1+\delta /2+\delta \chi _{k} \right)\varepsilon _{S} +\delta \left(1/2-\chi _{k} \right)\varepsilon _{1} }
{\left(1+\delta /2+\delta \chi _{k} \right)\sigma _{S} +\delta \left(1/2-\chi _{k} \right)\sigma _{1} } \approx T_{k1} .   
\end{equation} 
The main reason is as follows. At frequencies close to $1/T$ there is a huge change in 
$\tilde\epsilon_k$ of order $\varepsilon_{S} /(\delta (1/2-\chi_{k} ))$, 
and consequently a significant change in the total permittivity $\epsilon_{\mbox{\tiny sus}}$ 
given by \eqref{eq30}. Therefore $T$ provides an approximate value for the relaxation time $T_{k1}$ of the suspension effective permittivity 
$\epsilon_{\mbox{\tiny sus}}$.   

For a non-conductive shell $\sigma_{S}\approx 0$, 
or more precisely when $\sigma _{S} {\rm \ll }\delta \left(1/2-\chi _{k} \right)\sigma _{1}$,
the relaxation time \eqref{eq44} is:
\begin{equation} \label{eq45} 
T_{k1} \approx \; \varepsilon_{vac} \; \varepsilon_{S} /(\delta \cdot \sigma _{1} (1/2-\chi_{k} )) 
\end{equation} 
showing a strong dependence on the thickness of the shell  and on the shape of the particle, through the
eigenvalue $\chi_k$.
Due to the small parameter $\delta$ in \eqref{eq45} the first relaxation (i. e., membrane relaxation) 
tends to have a lower frequency than the second relaxation, 
which is present even for particles with no shell (see the discussion below). 
In addition, cumbersome but straightforward calculations provide the dielectric decrement $\Delta \varepsilon_{k1} $ in \eqref{eq42} 
\begin{equation} \label{eq46} 
\Delta \varepsilon_{k1} \approx 4 f p_k \varepsilon_{S} (\delta (1/2+\chi _{k} )^{2} (1/2-\chi _{k} ))^{-1} 
\end{equation} 
that is very large due to the same strong dependence on the thickness of the shell.
The effect is even more dramatic when 
the second largest eigenvalue is very close to the largest eigenvalue, 
$\left(1/2-\chi _{2} \right)\to 0$, like in the case of two cells connected by 
tight junctions. 

For a suspension of shelled spheres $\eta = 1 + \Delta R/R$ and $\delta =\eta ^{3} -1 \approx 3 \Delta R/R$, 
where $\Delta R$ is the 
thickness of the membrane, $R$ is inner radius, and $R + \Delta R$ is the total radius. Thus both 
$T_{k1}$ and $\Delta \varepsilon_{k1} $ are proportional to $R$ and $\varepsilon_{S}$ and inverse proportional 
to $\Delta R$ like in the Pauly-Schwan theory \cite{Pauly1959,Schwan1996}. 
Moreover, the dielectric decrement $\Delta \varepsilon_{k1} $ in \eqref{eq46} is 
a generalization of equation (54a) in Ref. \cite{Schwan1996}. In the same time, 
the relaxation time \eqref{eq45} differs with respect to equation (56a) in Ref. \cite{Schwan1996} 
only by the conductivity term. We will show elsewhere that a more appropriate treatment of the 
relaxation times recovers also the relaxation time given by equation (54a) in Ref. \cite{Schwan1996}.  

Thus, a non-conductive and thin shell/membrane produces a large relaxation of the complex
permittivity of the suspension \cite{Fricke1953}. The experimental evidences further 
support these theoretical facts:
when attacking the membrane with a membrane disrupting compound ( for example a detergent)
the relaxation almost vanishes as the cellular membrane is permeated \cite{Asami1977}. 

For frequencies higher than $1/T_{k1}$ the cell permittivity is essentially determined by the 
dielectric properties of the cytoplasm, and does not depend on membrane's properties. The second Debye 
relaxation occurs at higher frequencies than the first (membrane) relaxation,
and has the relaxation time 
\begin{equation} \label{eq48} 
T_{k2} \approx \varepsilon_{vac} \frac{\left(1/2+\chi _{k} \right)\varepsilon _{0} +\left(1/2-\chi _{k} \right)\varepsilon _{1} }
{\left(1/2+\chi _{k} \right)\sigma _{0} +\left(1/2-\chi _{k} \right)\sigma _{1} }
\end{equation}
derived from the pole of equation  \eqref{eq30}. 
The corresponding dielectric decrement is
\begin{eqnarray*}
&\Delta \varepsilon _{k2} \approx 
f p_{k} (1/2-\chi_k)(\varepsilon_1 \sigma_0 - \varepsilon_0\sigma_1)^2 \times\\
&\left((1/2+\chi_k)\varepsilon_0 +(1/2-\chi_k)\varepsilon_1\right)^{-1} \times\\ 
&\left((1/2+\chi_k)\sigma_0      +(1/2-\chi_k)\sigma_1\right)^{-2}   
\end{eqnarray*} 
The last two  equations are similar to the ones that are given for spherical particles in Ref. \cite{Schwan1996} 
(equations (46) and (49) in the aforementioned reference).
The relaxation given by $T_{k2}$ is basically the relaxation of a homogenous particle embedded in a dielectric
environment and was also discussed in Ref. \cite{Lei2001} by a closely related spectral method. 
If the conductivity of the cytoplasm is comparable to the conductivity 
of the outer medium, the decrement of the second relaxation is small such that it cannot be distinguished
in the spectrum.
On the contrary, if the conductivity of the outer medium is much greater or smaller that that of cytoplasm,
than a second observable relaxation occurs. 
Unlike the membrane relaxation, this second relaxation depends only weakly on the shape. 
By assuming that $\sigma _{0} \ll \sigma_1$ and by using
a finite-difference method, this resonance was also obtained in \cite{Asami2006}
and it was instrumental in explaining the 
experimental data on the fission of yeast cells of Asami \textit{et al.} \cite{Asami1999}
by Lei \textit{et al.} \cite{Lei2001}.

The shape of the particle is important because it affects the number of dipole-active eigenvalues and their
strengths. In principle, each dipole-active eigenvalue introduces a new relaxation in the dielectric spectrum,
providing this relaxation is well separated from the others. A cluster with complex geometry can have several
dipole-active eigenvalues, but unless the cluster is larger in one dimension then in the others, or there are
tight junctions, the relaxations overlap to create broad features in the spectrum. An extra relaxation is
introduced when the particles are covered by thin membranes. In addition, if $\left(1/2-\chi _{k} \right)\to 0$
for that eigenvalue, then the shell induced relaxation has low frequency, large relaxation time and 
large dielectric decrement. Based on the spectral BIE method, it is therefore possible to explicitly relate the dielectric
spectra of cell suspensions to cell's geometry and electric parameters, and, even design fitting procedures to
evaluate these parameters from measurements. 

\section{Results}

\subsection{Numerical procedure}

The calculation of the effective permittivity for a suspension uses equations \eqref{eq2} or \eqref{eq30},
and reduces then to finding the eigenvalues $\chi_k$ and eigenvectors $|u_k\rangle$ and $|v_k\rangle$ of
the linear response operator $M$. This problem is solved by employing a finite basis of {\em NB} functions
defined on the surface $\Sigma$. A natural basis for a surface not far from a sphere is the
generalized hyperspherical harmonics functions
\begin{equation} \label{eq31} 
\tilde{Y}_{lm} ({\bf x})=\frac{1}{\sqrt{s({\bf x})}}\;Y_{lm} \left(\theta({\bf x}),\varphi ({\bf x})\right), 
\end{equation}
where $s({\bf x})$ is related to the surface element through 
$d\Sigma = s({\bf x}) \; d\Omega_{\bf x} $
and $d\Omega_{\bf x} $ is the  solid angle element.

Another choice could be based on Chebyshev polynomials of the first kind \cite{Abramowitz1972}
\begin{equation} \label{eq32} 
\tilde{T}_{lm} ({\bf x})=\frac 1{\sqrt{s({\bf x})}} \; T_l\left(\theta ({\bf x})\right)\; 
e^{i m \varphi ({\bf x})} . 
\end{equation} 
Both bases are complete and orthogonal in the Hilbert space of square integrable functions defined on $\Sigma$.

In this paper, we model the linear cluster of particles as an object with axial symmetry. We seek to find a 
surface of revolution for which the thickness of the interparticle joints can be varied without perturbing the
overall shape of the object. We use two representations for the surface $\Sigma$: 
(A) for clusters of two
particles we use spherical coordinates 
$\{x,y,z\} = \{r(\theta)\sin\theta \cos\phi, \; r(\theta)\sin\theta \sin\phi, \; r(\theta) \cos\theta\}$,
and (B) for clusters with more
than two particles we specify the surface in terms of a function $g(z)$ as
$\{x,y,z\} = \{g(z)\cos\phi,\; g(z)\sin\phi, \;z\}$.

In the case B, the surface element is
\begin{equation} \label{eq35} 
d\Sigma = g(z) \sqrt{1+g'^2(z)} \; dzd\varphi,
\end{equation} 
and the normal to surface $\Sigma $ is
\begin{equation} \label{eq36} 
{\bf n}=\frac 1{\sqrt{1+g'^2(z)}}\; 
\left(\begin{array}{c} \cos\varphi \\ \sin\varphi \\ -g'(z) \end{array}\right). 
\end{equation} 

In the basis of generalized hyperspherical harmonics the operator $M$ has matrix elements
given by
\begin{widetext}
\begin{equation} \label{eq38} 
M_{lm; l'm'} = \delta_{mm'} 
\iint\limits_0^{\kern 10pt 2\pi} \;\iint\limits_{z_{\mbox{\tiny min}}}^{\kern 10pt z_{\mbox{\tiny max}}}
A(z,z', \varphi - \varphi')
P_l^m(\cos\theta(z)) P_{l'}^{m'}(\cos\theta(z')) \; e^{im(\varphi - \varphi')}
\; G(z, z')
\; dz\;dz'\;d\varphi\;d\varphi'
\end{equation} 
where
\begin{equation}
G(z,z') = \sqrt{g(z) g(z') \sqrt{(1+g'(z)) (1+g'(z'))^{-1}}} ,
\end{equation}
and
\begin{equation}
A(z,z',\phi) =  \frac{(g(z) - g(z')) \cos\phi - (z-z') g'(z)}
{\left[ g^2(z) +g^2(z') - 2 g(z) g(z') \cos\phi + (z-z')^2\right]^{3/2}}
\end{equation} 
\end{widetext}

After the angle integration in equation \eqref{eq38} and by using the elliptic
integrals given in the Appendix, the matrix elements are obtained by numerical
evaluation of the resulting ($z$, $z'$) double integral using an {\em NQ}-point
Gauss-Legendre quadrature \cite{Boyd2001,Abramowitz1972}. Because of the integrable singularity apparent in the
kernel of the operator $M$ in equation \eqref{eq6}, the mesh of $z$ must be shifted from the
mesh of $z'$ by a transformation which insures that there is no overlap between the
two meshes.

The delta symbols $\delta_{mm'}$ in equation \eqref{eq38} reflects the fact that we consider only objects with
rotational symmetry in this paper. Moreover, for fields parallel with the cluster axis $m=0$, while
$m=1$ for perpendicular fields.

The convergence of the results is obtained in two steps. First, the number {\em NQ} of quadrature
points is increased until the matrix elements of $M$ converge, and then the size {\em NB} of the
basis set is increased until the relevant eigenvalues $\chi_k$ and their corresponding weights $p_k$
have acquired the desired accuracy. A necessary test for convergence is the fulfillment of the
sum rules $\sum_k P_{k,i} = 1$ and $\sum_{i,k} \chi _k P_{k,i} = 1/2$ with sufficient accuracy \cite{Bergman1978,Bergman1992}.
Usually the convergence is fast in both the number of quadrature points and the size of basis,
unless the system has a tight junction where some care must be considered in order
to achieve the required accuracy of the eigenmodes with the eigenvalues close to $1/2$.
For a sphere there is just one dipole-active eigenmode which has eigenvalue
$\chi =1/6$ and weight $p=1$, while for an ellipsoid there is one dipole-active eigenmode 
along each axis.
Fast and accurate solutions are achieved for spheroids with a basis size of {\em NB} =20 and
with {\em NQ} = 64 quadrature points. In general, the size of the basis and the 
number of the quadrature points increase with the number of cells in the cluster and with the decreasing of the junction size.
Thus, for our numerical examples a basis with {\em NB}=35-40 and {\em NQ} = 128 quadrature points are enough
for a converged solution in the case of the dimers and {\em NB}=50 and {\em NQ} = 200 quadrature points in the case 
of the clusters with up to four cells.

\subsection{Two cells joined by tight junction}

The equation $r(\theta) = (h + \cos^2\theta)/(1 - a\cos^2\theta)$
describes the shape of a two-particle cluster. Parameter $h$ controls the tightness of the inter-particle
junction and parameter $a$ measures the deviation from a spherical shape.
More precisely, $h$ is the radius of the smallest circle at around the thinnest part of the junction.

\begin{figure}[h]
\includegraphics [width=5.0in] {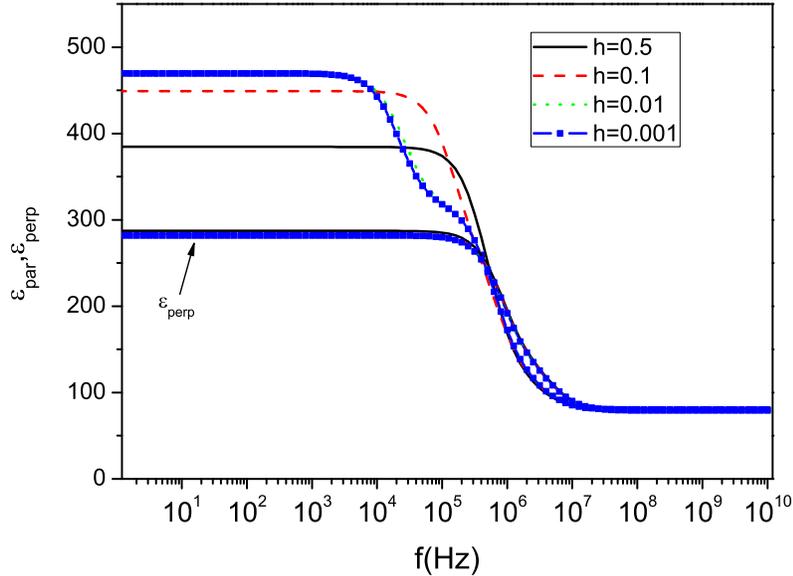}
\caption{\label{fig:1}
(Color online) The spectrum of the
effective permittivity of a suspension of dimers with various junction thickness $h$, and
with parallel and perpendicular field configurations.
The suspension permittivity for an electric field perpendicular to cluster axis does not depend
on $h$ and is pointed by an arrow.
}
\end{figure}

Figure \ref{fig:1} shows the effective permittivity for a suspension of particles with the following parameters:
$\varepsilon_1 = 70$, 
$\sigma _{1}   = 0.25$ S/m,
$\varepsilon_S = 6$,
$\sigma_S = 0$,
$\varepsilon_0 = 81$,
$\sigma_0 = 0.374$ S/m,
volume fraction $f=0.05$, membrane thickness $\delta = 0.00947275$, and $a=0.2$.
The effective permittivity does not depend on the thickness parameter $h$ when the stimulus electric field is perpendicular to
the cluster axis. However, a new relaxation becomes apparent as $h \to 0$,
for parallel fields \cite{Asami2006, Gheorghiu2002}. 

\begin{figure}
\includegraphics [width=4.0in] {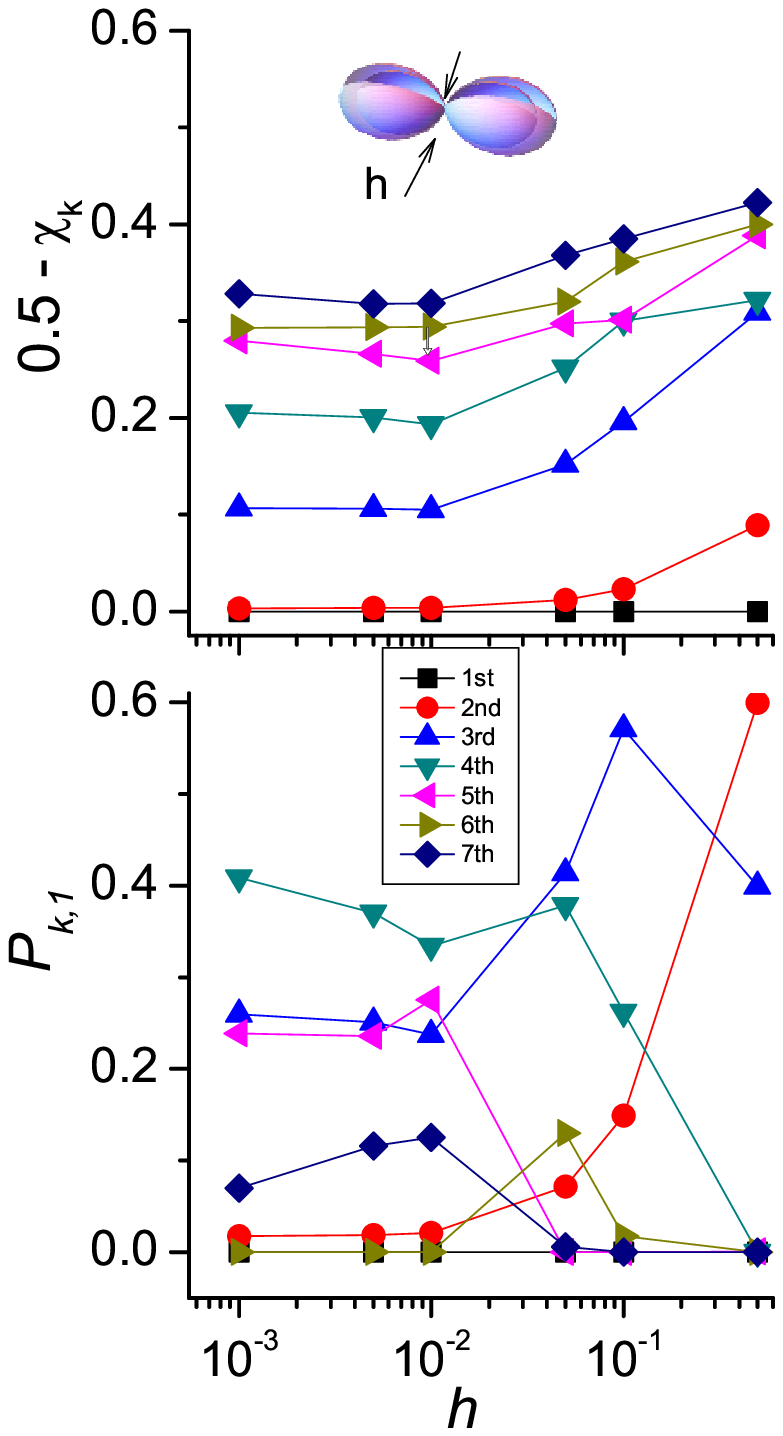}
\caption{\label{fig:2}
(Color online) The largest seven eigenvalues and their weights for a binary cluster
with parameter $a=0.2$, as a function of $h$. The inset shows the shape of the dimer. 
}
\end{figure}

Figure \ref{fig:2} presents the first 7 eigenvalues $\chi _{k} $ and their weights $P_{k,1} $ for a field parallel to the $z$ axis. 
As the junction become tighter ($h\to 0$) more eigenvalues become dipole-active. While all eigenvalues are important in shaping the 
dielectric spectrum, the second largest eigenvalue $\chi _{2} $ is crucial to explaining the occurrence of an additional relaxation at 
low frequencies, as observed for small $h$ in \cite{Asami2006, Gheorghiu2002}.
Although its weight $P_{2,1} $ also decreases for small $h$, this dipole active 
eigenmode approaches 1/2 as the junction becomes tighter. 
Thus, according to equations \eqref{eq45} and \eqref{eq46} the effect of $\chi _{2} $ is 
``enhanced'' due to the presence of a nonconductive shell 
(as the case for biological particles analyzed in \cite{Asami2006, Gheorghiu2002}). Moreover, 
the decrease of $P_{2,1} $ is compensated by the increase of $1/(1/2-\chi _{2} )$.

\begin{figure}
\includegraphics[width=4.0in]{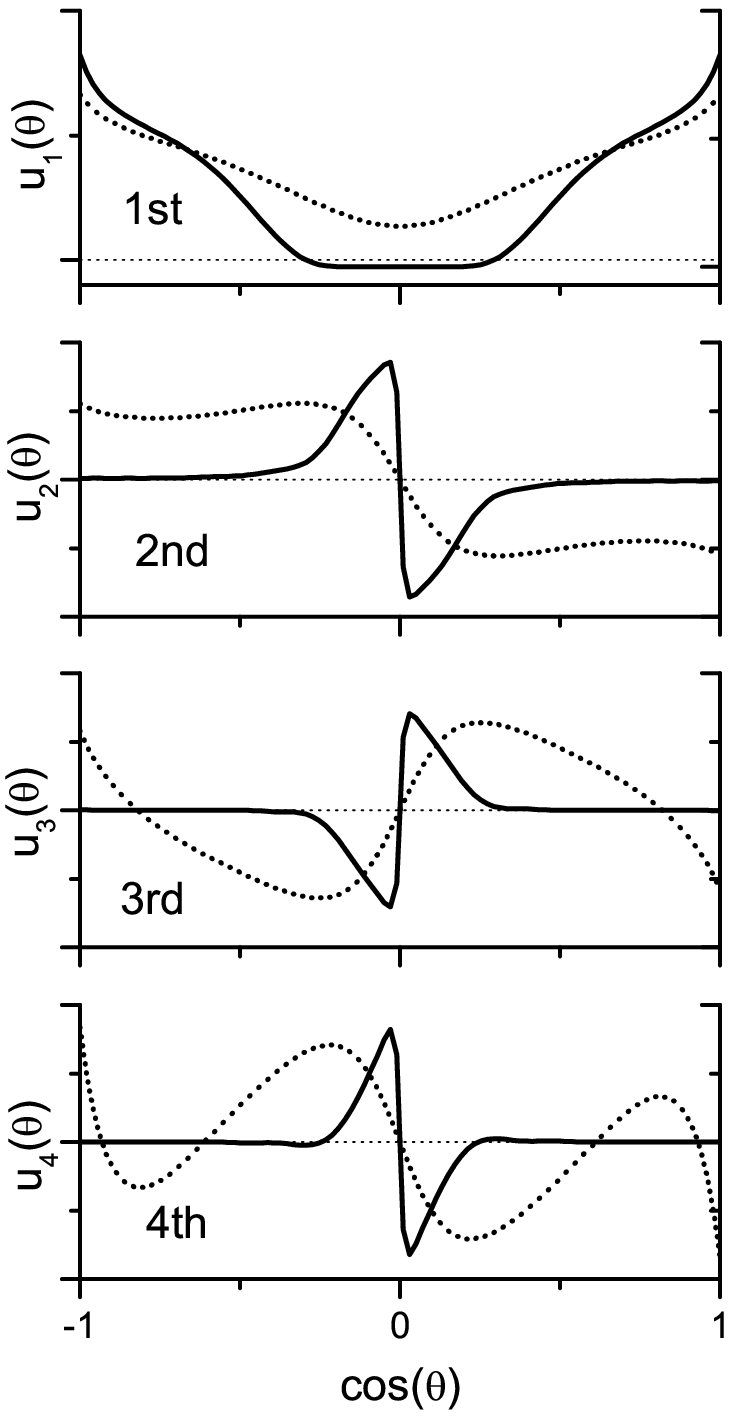}
\caption{\label{fig:3}
The first four eigenvectors for a dimer given by equation $r(\theta )$ ; $a=0.2$ and $h=0.5$ (dotted line) and $h=0.01$ (solid line). 
}
\end{figure} 

The presence of a new relaxation at low frequency along with its relationship with the size of $h$ has been already singled out in 
Gheorghiu \textit{et al.} \cite{Gheorghiu2002} by using the same method but without the analysis of dipole-active eigenmodes. 
Using a finite discrete model \cite{Asami2006}, the relaxation was observed before the segregation during cell division,
while other papers \cite{Biasio2007,Biasio2009} fail to relate 
the size of $h$ to the new relaxation, even though one of them \cite{Biasio2007}
employs essentially the same method as the one outlined in the present work.

Figure \ref{fig:3} shows the charge distribution associated with the first four eigenvalues for two distinct values of $h$.
The second eigenmode is an antisymmetric combination of net charge distributions (monopoles) on each particle of the dimer.
The third charge distribution is an antisymmetric 
combination of charge distributions with a dipole moment on each part of the dimer and
the forth distribution is antisymmetric combination of charge distributions with a quadrupole moment on each particle.
At small $h$ (tight junctions), charge accumulates in the vicinity of the junction \cite{Klimov2007}.

\subsection{Clusters of more than two particles}

\begin{figure}
\includegraphics [width=4.0in]{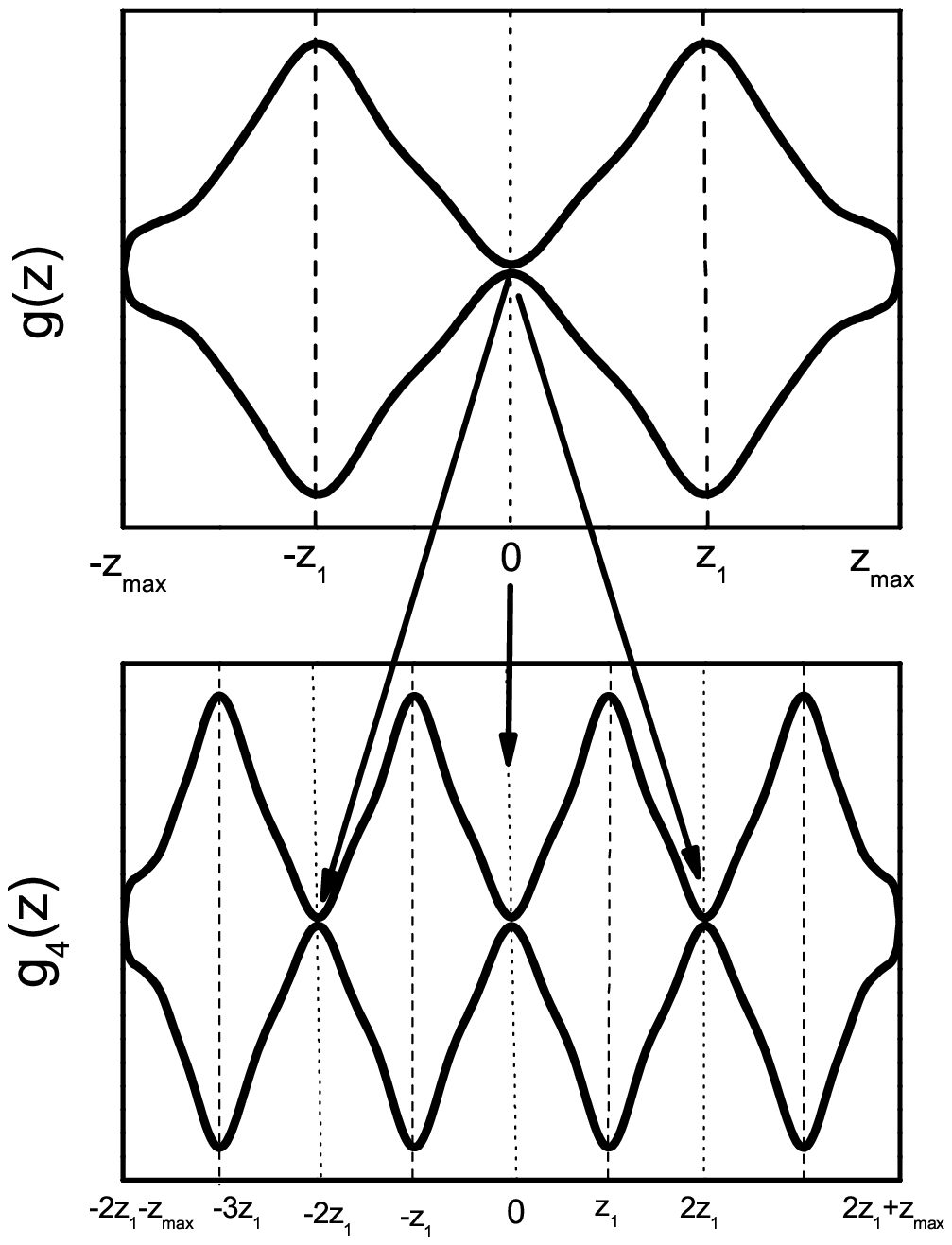}
\caption{\label{fig:4}
Smooth construction of a cluster (lower panel) from a dimer (upper panel). 
The parts determined by $z\in \left[-z_{1} ,z_{1} \right]$ are ``glued'' together with the ends of the dimer. 
The arrows show where the junctions will be placed in the cluster. 
}
\end{figure}

Smoother yet tight junctions would bring $\left(1/2-\chi _{2} \right)$ closer to 0 than sharp and tight junctions. The reason is simple: smoother 
junctions have the two parts of the dimer farther apart.
We have analyzed linear clusters of cells connected by smooth and tight junctions by using a
($z$, $\phi$) parameterization which describes
a surface by  $\{ x=g(z)\cos \varphi ,y=g(z)\sin \varphi, z\}$. 
The construction starts from a dimer shape that resembles the shape of
the epithelial cells like MDCK (Madin-Darby Canine Kidney) cells. An example of such shape, displayed in
figure \ref{fig:4}, extends
from $-z_{\mbox{\tiny max}}$ to $z_{\mbox{\tiny max}}$ and it can be decomposed in three parts: 
the left cap ($-z_{\mbox{\tiny max}} \le z \le -z_1$),
the central part ($ -z_1 \le z \le z_1$), and the right cap ($z_1 \le z \le z_{\mbox{\tiny max}}$).
At position $\pm z_1$ the shape function
has its maximum. An $m$-cell linear cluster is obtained by repeating the central part
$m-1$ times and it extends from $-L_m$ to $L_m$,
where $L_m = z_{\mbox{\tiny max}} + (m-2) z_1$. Mathematically, the shape is described by:
\begin{widetext}
\begin{equation} \label{eq50} 
g_m(z)=\left\{
\begin{array}{l} 
g(z+(m-2)z_{1} ), \mbox{\quad for}\; -L_m \le z \le -L_m + z_{\mbox{\tiny max}} \\ 
g\left(\mbox{mod}(z+(1+(-1)^{m})z_1/2, 2z_1) - z_1\right), \mbox{\quad for}\;
-L_m + z_{\mbox{\tiny max}} \le z \le L_m - z_{\mbox{\tiny max}}\\ 
g(z-(m-2)z_1), \mbox{\quad for}\; L_m \le z \le L_m - z_{\mbox{\tiny max}},
\end{array}\right.  
\end{equation} 
\end{widetext}
where mod($x$, $y$) is the remainder of the division of $x$ by $y$. For the examples considered here, the dimer shape function is:
\begin{equation} \label{eq51} 
\begin{array}{l} 
g(z)=0.01+2.32317\;z^2 -11.9862\;z^4 +40.4045\;z^6\\
-74.2226\;z^8 +79.142\;z^{10} -51.8929\;z^{12} +21.3096{\rm \; }z^{14}\\
-5.35113\;z^{16} +0.752147\;z^{18} -0.045375\;z^{20} ,
\end{array} 
\end{equation} 
with $z_{\mbox{\tiny max}} = 1.77377$ and $z_1 = z_{\mbox{\tiny max}}/2$.

Tables \ref{para} and \ref{perp} list the most representative dipole-active eigenmodes for a trimer in perpendicular and parallel fields.
Only the parallel field configuration has a dipole-active eigenvalue close to 1/2, with a relatively small weight.
\begin{table}
\caption{Most representative dipole-active eigenmodes and their weights for the trimer in parallel field.
}
\label{para}
\begin{ruledtabular}
\begin{tabular}{lcccccc}
   & $\chi _{k}$ &   $P_{k,1} $  \\
\hline
&   0.4996  &    0.01305 \\
&   0.40642 &    0.07769 \\
&   0.40448 &    0.1391  \\
&   0.37763 &    0.17366 \\
&   0.26409 &    0.20519 \\
&   0.23809 &    0.01839 \\
&   0.17557 &    0.05104 \\
&   0.13522 &    0.05234 \\
&   0.12165 &    0.01066 \\
&   0.07192 &    0.06197 \\
&   0.06649 &    0.01239 \\
&   0.03479 &    0.03899 \\
&   0.03362 &    0.02823 \\
&   0.0281  &    0.04688 \\
&   0.02377 &    0.02519 \\
\end{tabular}
\end{ruledtabular}
\end{table}

\begin{table}
\caption{Most representative dipole-active eigenmodes and their weights for the trimer in perpendicular field.
}
\label{perp}
\begin{ruledtabular}
\begin{tabular}{lcccccc}
   & $\chi _{k}$ &   $P_{k,2} $  \\
\hline
&    0.1831   &     0.64102 \\
&    0.06964  &     0.01464 \\ 
&    0.05492  &     0.01343 \\ 
&    0.0272   &     0.01065 \\ 
&    0.01674  &     0.05687 \\ 
&    0.01479  &     0.13179 \\ 
&    0.00395  &     0.05029 \\ 
&    -0.01635 &     0.03246 \\ 
\end{tabular}
\end{ruledtabular}
\end{table}

\begin{figure}
\includegraphics [width=5.0in] {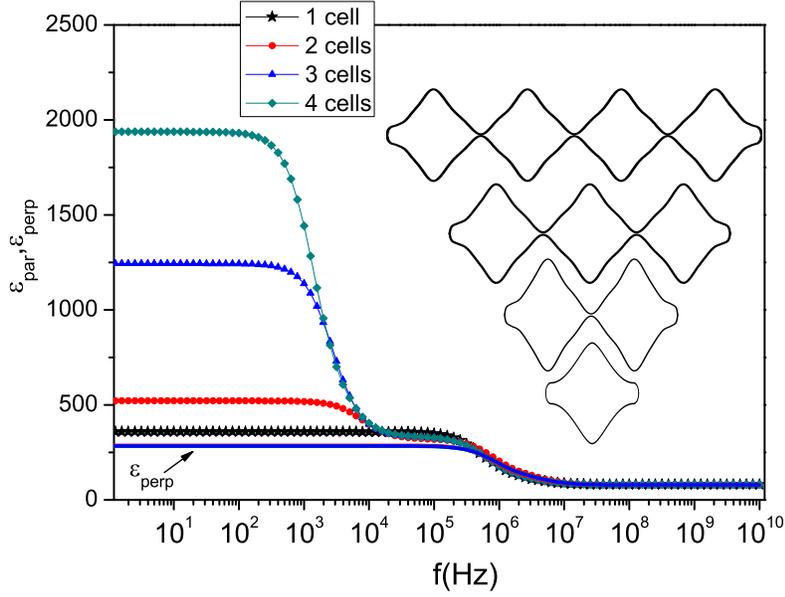}
\caption{\label{fig:5} 
(Color online) Effective permittivity for clusters (shown in the inset) of one, two, three, and four cells connected by tight and smooth junctions. 
The field is either parallel (solid lines with symbols) or perpendicular (solid lines only) to the cluster axis. The effective 
permittivity either increases strongly with the number of cells for parallel geometry, or does not change for a perpendicular geometry. 
}
\end{figure}

The results for linear clusters of up to four particles are displayed in Figure \ref{fig:5}.
The electric parameters are the same as ones used for dimers in the previous section.
An additional, distinct low-frequency relaxation emerges for clusters with more then one particles,
only when the stimulus field is parallel with the symmetry axis. 
The relaxation frequency decreases, while the intensity of these relaxations increases, as the number of cluster members increases.
This behavior is explained again by the combination of eigenvalues close to 1/2, with thin non-conductive 
layers covering the cluster and is consistent 
with experimental data on ischemic tissues \cite{Knapp2005}, which reports that the cell separation (closure of gap-junctions)
is responsible for decrease and eventual disappearance of the low-frequency dispersion.
 
\begin{figure}
\includegraphics [width=5.0in] {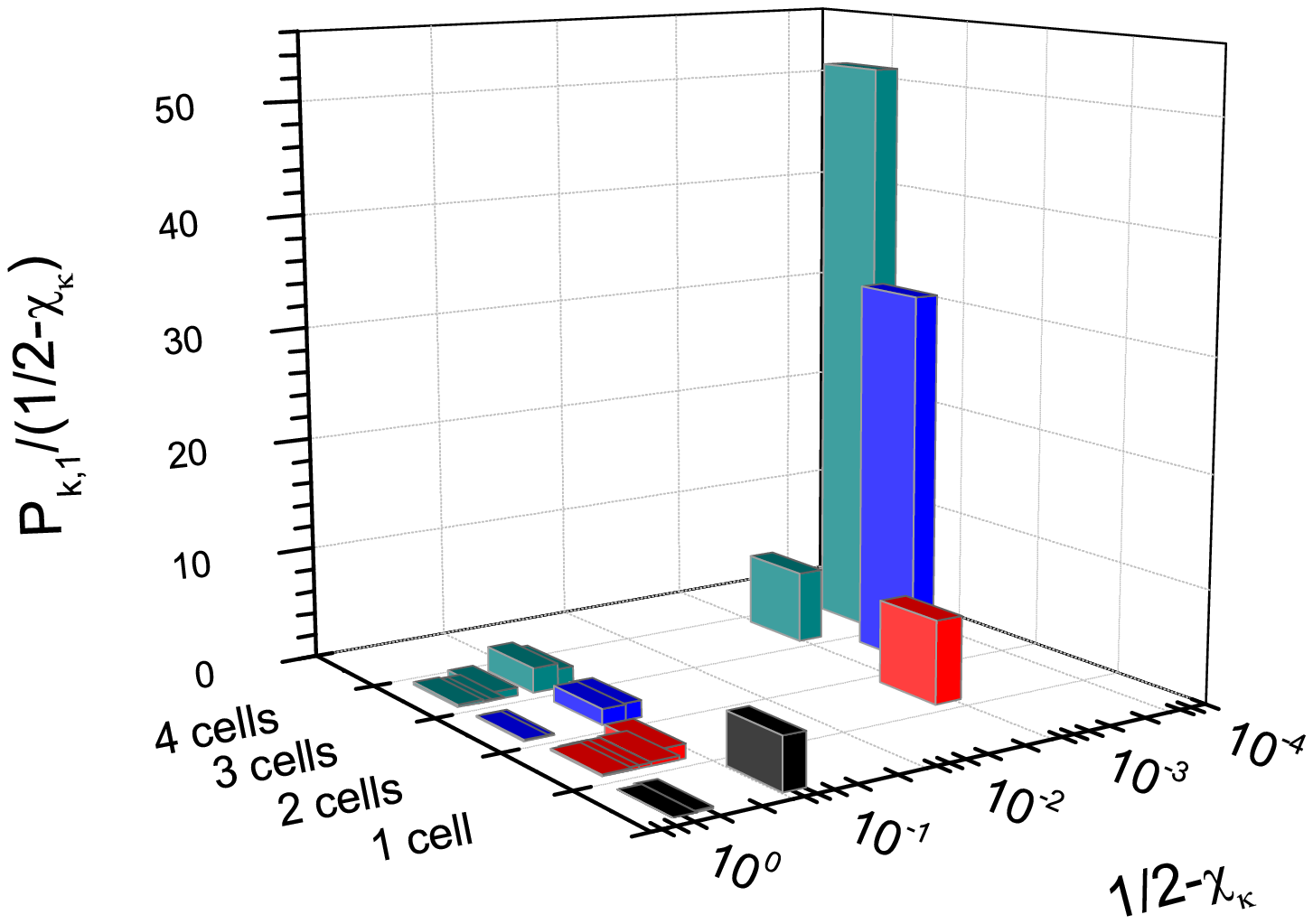}
\caption{\label{fig:6}
(Color online) $P_{k,1} /(1/2-\chi _{k} )$ versus $(1/2-\chi _{k} )$ for clusters of up to four cells. 
The second eigenvalue has the largest contribution to intensity of relaxation. 
}
\end{figure}

In Figure \ref{fig:6} we plot $P_{k,1} /(1/2-\chi _{k} )$ versus $(1/2-\chi _{k} )$, which shows that the number of 
dipole-active eigenmodes increases with the number of particles in the cluster.
According to \eqref{eq45} and \eqref{eq46}, Figure \ref{fig:6} shows in fact the dielectric 
decrement versus its corresponding relaxation frequency for each dipole-active eigenmode of the given clusters.
For clusters of two or three particles, there is one 
important active eigenmode close to $1/2$, while for clusters of four particles there are two active eigenmodes.

\begin{figure}
\includegraphics [width=4.0in] {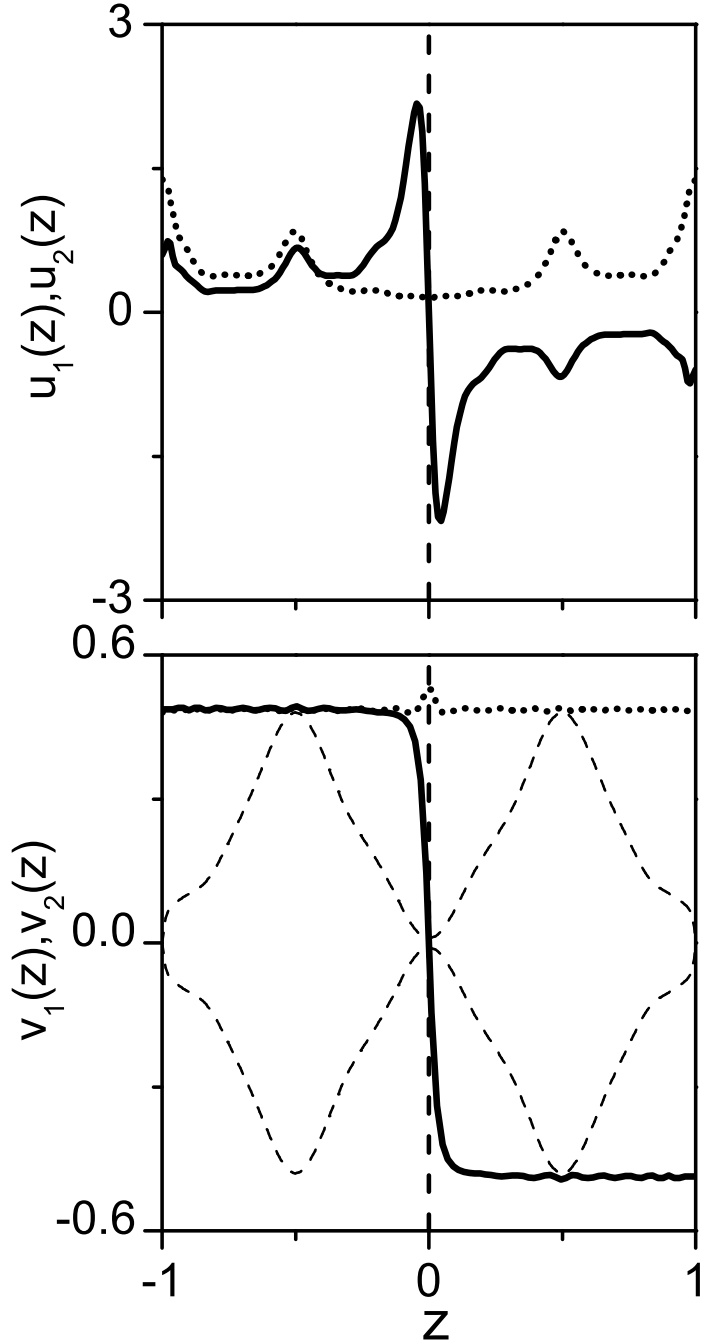}
\caption{\label{fig:7}
The first (dotted line) and the second (solid line) eigenfunction of $M$ (upper panel) and $M^{\dag } $ (lower panel) for a dimer whose shape is 
depicted by dashed line in the lower panel. The second eigenfunction of $M^{\dag } $ is an antisymmetric combination of almost constant 
distributions on each part of the dimer. 
}
\end{figure}

It can be conjectured that for a general linear cluster made of $m$ particles,
there are $m-1$ eigenvalues close to $1/2$, of which 
the largest one is always dipole-active and has the largest weight. 
In fact one can show that for two cells connected by smooth and tight 
junction characterized by parameter $h$, $\left(1/2-\chi _{2} \right)\propto h^{2} $ 
when $h\to 0$, or more precisely $\left(1/2-\chi _{2} \right)$ is 
proportional with the solid angle encompassed by the missing part of a cell
when it is connected with other cell in the dimer. The proof is based on 
the theorem of the solid angle \cite{Vladimirov1984}. The generalization to a finite cluster is 
also straightforward to $\left(1/2-\chi _{2} \right)\propto h^{2} /m$ (in that case the solid 
angle encompassed by the middle junction is proportional to $h^{2} /m$.
The weight of the second eigenmode is
$P_{2,1} = \langle {\bf x} \cdot {\bf {N}_{1}}|u_{2}\rangle \langle v_{2} |{\bf n} \cdot {\bf {N}_{1}}\rangle /V_{1}$. 
If we consider that the surface of the cluster is determined by the function $g(z)$ then, up to a constant 
factor, $\langle v_{2} |{\bf n} \cdot {\bf {N}_{1}}\rangle \approx g(0)^2 = h^2$ for two cells connected by 
smooth and tight junctions. The proof considers that the second eigenfunction of $M^{\dag }$ is an antisymmetric combination 
of constant distributions on each part of the dimer. This assertion is confirmed in Figure \ref{fig:7}. 
Moreover, $\langle {\bf x} \cdot {\bf {N}_{1}}|u_{2}\rangle /V_{1} $ is weakly dependent on $h$.
Therefore, for a parallel setting of the field stimulus, 
$P_{2,1} /(1/2-\chi _{2} )$, which is the measure of the dielectric decrement of low-frequency relaxation,
is finite and it increases when the number of cells is increased.
The increase of relaxation decrement when $m\to \infty $ is physically limited by 
$\sigma _{S} {\rm \ll }\delta \left(1/2-\chi _{k} \right)\sigma _{1} $, since the membrane conductivity is not strictly 0.

Due to cluster's shape and membrane properties, the variation of $P_{k,1} /(1/2-\chi _{k} )$ and $(1/2-\chi _{k} )$
with respect the eigenmode $k$ 
determines a low frequency relaxation when the dipole-active eigenvalue $\chi _{k}$ is close to 1/2. 
We note here that for dipole-active eigenmodes of ellipsoids the 
term $(1/2-\chi _{k} )$ is called the depolarization factor 
and has analytical expression \cite{Venermo2007}.
Prolate spheroids with longitudinal axis much larger than the transverse axis (needles) have the 
longitudinal depolarization factor approaching 0 and the transverse depolarization factor approaching 1/2. 
More precisely, for a long prolate spheroid, the longitudinal depolarization factor scales as 
$(1/2-\chi _{2} )\propto a_x^2/a_z^2,\; a_z > a_x = a_y$, as $a_x \to 0$.
On the other hand, extensive numerical calculations support the fact that cylinders with the same aspect ratio behave similarly 
to prolate spheroids \cite{Venermo2007}.
Thus, it is not hard to observe that the low-frequency relaxation of linear clusters of cells connected
by tight junctions is similar to that of a needle or a thin cylinder as long as the cluster and as thick as the junction.
On the other hand, the high-frequency relaxation of the cluster shows
the relaxation of a suspension of spheroids with the same volume as the volume of 
a single cell.  Therefore the dielectric spectrum for a suspension of clusters is the same as the spectrum
of a two species suspension made of thin cylinders and spheroids.

\section{Conclusions}

We present a theoretical framework based on a spectral representation of BIE 
and able to calculate the dielectric behavior 
of linear clusters with a wide range of shapes and dielectric structures.
The theory agrees with the results of Pauly and Schwan for a sphere covered with a shell \cite{Pauly1959,Schwan1996}. 
In fact, for spheroids, our theory is the same as the analytical results of Asami \textit{et al.} \cite{Asami1980}. 
We present extensive calculations of clusters with shapes resembling MDCK cells. 

A practical numerical recipe to compute the effective permittivity of linear clusters with arbitrary number of cells is provided.
Examples are given for cluster with shapes described as $r(\theta)$ in spherical coordinates or
using ($z$, $\varphi$) parameters as $\{ x=g(z)\cos \varphi ,y=g(z)\sin \varphi, z \}$.
Other studies in the literature used only spherical coordinates representation \cite{Gheorghiu2002,Prodan1999,Biasio2009}.
A direct relation between the geometry and dielectric parameters of the cells and 
their dielectric behavior described by a Debye representation has been formulated for the first time. 
Other work \cite{Lei2001}, which is based on a closely related spectral method 
\cite{Bergman1978,Bergman1992,Stockman2001,Li2003}, 
found a direct relation linking the geometry
and electric parameters to the dielectric behavior only for homogenous particles.
Moreover, the method used in \cite{Lei2001} treats only particles with spheroidal geometry. 

We show that the spectral representation provides a straightforward evaluation of
the characteristic time constants and dielectric decrements 
of the relaxations induced by cell membrane. 
We prove that the effective permittivity is sensitive to the shape of the embedded particles, 
specially when the linear response operator has strong dipole active modes (with large weights $p_{k} $).
A low-frequency and distinct relaxation occurs 
when the largest dipole-active eigenvalue is very close to $1/2$. 
Clusters of living cells connected by tight junctions or very long cells have 
such an eigenvalue.
Our results also shed a new light on the understanding of recent numerical calculations \cite{Ron2009}
performed with a boundary element method on clustered cells where the low-frequency relaxation is 
attributed to the tight (gap) junctions connecting the cells.
The method used in \cite{Ron2009} does not use the confocal geometry assumption .

The present work has several implications and applications.
We emphasize the capabilities of dielectric spectroscopy to monitor the 
dynamics of cellular systems, e.g., cells during cell cycle division,
using synchronized yeast cells \cite{Gheorghiu1998,Asami2006,Knapp2005}, 
or monolayers of interconnected cells \cite{Wegener2000,Urdapilleta2006}.
Also the method is able to assess the dielectric behavior of linear aggregates or rouleaux of erythrocytes, 
where the ellipsoidal or cylindrical approximations are not adequate \cite{Sebastian2005,Asami2007}. 

The proposed representation is a powerful alternative to finite element or other purely numerical approaches,
because  it provides the analytical framework to explain 
and predict the complex dielectric spectra occurring in bioengineering applications.
Extension of this method to other surfaces of revolution, for example linear clusters with more
than 4 particles, is straightforward providing an adequate parametric equation is available.
Finally, in many cases (e.g. shapes with high symmetry) 
the method is faster, offers accurate solutions and last but not least can be integrated in
fitting procedures to analyze experimental spectra.

\begin{acknowledgments}
This work has been supported by Romanian Project ``Ideas'' No.120/2007 and FP 7 Nanomagma No.214107/2008. 
\end{acknowledgments}

\appendix*
\section{Integration over $\varphi$ }

The integrals over $\left(\varphi -\varphi '\right)$ are performed with the following elliptic integrals,

\begin{equation} \label{eq55} 
\int _{0}^{\pi } \frac{1}{\left(a-b\cos \varphi \right)^{3/2} } d\varphi =\frac{2}{\sqrt{a-b} } \frac{1}{a+b} E\left(-\frac{2b}{a-b} \right) 
\end{equation} 

\begin{widetext}
\begin{equation} \label{eq56} 
\int _{0}^{\pi } \frac{\cos \varphi }{\left(a-b\cos \varphi \right)^{3/2} } d\varphi =\frac{2}{\sqrt{a-b} } \frac{1}{b} \left[\frac{a}{a+b} E\left(-\frac{2b}{a-b} \right)-K\left(-\frac{2b}{a-b} \right)\right], 
\end{equation} 

\begin{equation} \label{eq57} 
\int _{0}^{\pi } \frac{\mathop{\cos }\nolimits^{2} \varphi }{\left(a-b\cos \varphi \right)^{3/2} } d\varphi =\frac{2}{\sqrt{a-b} } \frac{1}{b^{2} } \left[\frac{2a^{2} -b^{2} }{a+b} E\left(-\frac{2b}{a-b} \right)-2aK\left(-\frac{2b}{a-b} \right)\right], 
\end{equation} 
\end{widetext}
where $K(x)$ and $E(x)$ are the complete integrals of the first and second kind, respectively \cite{Abramowitz1972}.


\end{document}